\documentclass[12pt]{article}
\usepackage{axodraw}
\usepackage{epsfig}
\usepackage{amssymb}

\newcommand{\be}{\begin{equation}}
\newcommand{\ee}{\end{equation}}
\newcommand{\bea}{\begin{eqnarray}}
\newcommand{\eea}{\end{eqnarray}}

\newcommand{\ba}{\begin{array}}
\newcommand{\ea}{\end{array}}

\begin{document}

\centerline{\LARGE\bf Pion structure functions at 2 loops:}
\centerline{\LARGE\bf final state interaction with spectator}
\vspace{10mm}
\centerline{\large K.N. Zyablyuk}
\vspace{3mm}
\centerline{\tt zyablyuk@itep.ru}
\vspace{5mm}
\centerline{\it Institute of Theoretical and Experimental Physics,}
\centerline{\it B.Cheremushkinskaya 25, Moscow 117218, Russia}

\begin{abstract}
Pion structure functions are considered in the framework of QCD sum rules at the next to leading 
order. The 2-loop diargam, representing the gluon exchange between the struck quark and spectator,
is calculated for space-like pion momentum $p^2<0$.  After Borel transformation we obtain a leading twist 
correction to the pion structure function, which does not violate Bjorken scaling. Partial results
can be used in the calculation of complete set of  $\alpha_s$ corrections to the pion structure functions,
which may serve as an intitial condition in the evolution equation. Probabilistic interpretation
of the structure functions is discussed.
\end{abstract}

\section{Introduction}

The nucleon structure functions, namely, the parts of the 2-point electromagnetic current correlator
\be 
W_{\mu\nu}  \,=\,  {1\over 4\pi}  \int d^4 z \, e^{iqz} \, \langle N(p) | j_\mu^{em} (z) j_\nu^{em} (0) | N(p) \rangle 
\ee
are measured in the process of deep inelastic scattering. In the Bjorken limit they represent the 
sum of the probabilities $f_i(x)$ to find a parton $i$ carrying the part $x=Q^2/2(pq)$ 
of the nucleon momentum $p$. This long time known statement is based on the Wilson
Operator Product Expansion, which leads to the so-called twist expansion (for a review see \cite{J}):
\be
W_{\mu\nu} \sim \sum {C\over x^J}  \left({M\over Q}\right)^{d-J-2} 
\label{tw_ser}
\ee
where $Q^2=-q^2$, dimensionless constants $C$ originate from local operators $O_{\mu_1\ldots\mu_J}$
of dimension $d$ and spin $J$, averaged by the nucleon state:
\be
\langle N(p) | O_{\mu_1\ldots\mu_J} | N(p) \rangle \, =\, p_{\mu_1} \ldots p_{\mu_J} M^{d-J-2} C \, +\, \ldots
\label{oper_dj}
\ee
The most important are the lowest twist biquark operators $O\sim {\bar q}\gamma_\mu q$. Introducing the light-cone
wave functions, one comes to a probabilistic interpretation of the nucleon structure functions at high $Q^2$.

The leading order contribution of the 2-quark operator is commongly visualized by the following graph:
\be
\begin{picture}(150,55)
\Photon(20,45)(50,40)25
\Photon(120,45)(90,40)25
\Line(50,40)(90,40)
\Line(50,7)(50,40)
\Line(90,7)(90,40)
\Line(20,7)(50,7)
\Line(20,5)(120,5)
\Line(20,3)(120,3)
 \Line(120,7)(90,7)
\Text(11,10)[t]{$N$}
\Text(11,50)[t]{$\gamma^*$}
\end{picture}
\label{2q_diag}
\ee
The lower lines stand for the soft nucleon constituents (spectator), but the upper one shows hard
(struck) quark, interacting with the virtual photon. QCD interaction has several topologically distinct 
diagrams:
\be
\begin{picture}(300,70)
\Photon(0,65)(20,60)23
\Photon(80,65)(60,60)23
\Line(20,60)(60,60)
\Line(20,27)(20,60)
\Line(60,27)(60,60)
\Line(0,27)(20,27)
\Line(0,25)(80,25)
\Line(0,23)(80,23)
\Line(80,27)(60,27)
\Gluon(20,43)(60,43)36
\Text(41,15)[t]{a) evolution}
\Photon(130,65)(150,60)23
\Photon(210,65)(190,60)23
\Line(150,60)(190,60)
\Line(150,27)(150,60)
\Line(190,27)(190,60)
\Line(130,27)(150,27)
\Line(130,25)(210,25)
\Line(130,23)(210,23)
\Line(210,27)(190,27)
\Gluon(150,47)(174,25)35
\Text(171,20)[t]{b) initial state}
\Photon(260,65)(280,60)23
\Photon(340,65)(320,60)23
\Line(280,60)(320,60)
\Line(280,27)(280,60)
\Line(320,27)(320,60)
\Line(260,27)(280,27)
\Line(260,25)(340,25)
\Line(260,23)(340,23)
\Line(340,27)(320,27)
\Gluon(300,60)(300,25)35
\Text(301,20)[t]{c) final state}
\Text(240,7)[t]{interaction with spectator}
\end{picture}
\label{qcdint_diag}
\ee
The interaction of the struck quark with gluon field (\ref{qcdint_diag}a) leads to renormalization 
of the operators (\ref{oper_dj}). It yields the $\ln{Q^2}$ dependence of the structure functions,
governed by the evolution equation. Various anomalous dimensions have been calculated 
at the 2-loop order \cite{2loopEV}, and the 3-loop results were published recently \cite{3loopEV}.
The diagrams (\ref{qcdint_diag}b), (\ref{qcdint_diag}c) with spectator involved display the contribution
of the quark-quark-gluon operator. Consequently, they are either a gauge artefact \cite{S} or
higher twist corrections \cite{JMT}. They do not admit any probabilistic intepretation.

However, recently a statement appeared in literature \cite{BHMPS}, that initial or final state interaction
with spectator may give rise to a leading twist corrections, so the propabilistic interpretation of the
structure functions should be reconsidered.  In \cite{BHMPS} it was demonstrated by a model calculation of
the difractive dissociation cross section. A source of these corrections is the P-exponent between the
quark fields, necessary to keep the gauge invariance of the structure functions. Although
its argument vanishes in the light cone gauge $A^+=0$, it may generate a nonzero contribution
due to additional pole $1/k^+ $ of the gluon propagator here, as the authors of \cite{BHMPS} demonstrated.
For a more discussion of the P-exponent role see \cite{JY}. The initial and final state interactions 
were also shown \cite{Betal} to give a leading-twist contribution in Drell-Yan process and semi-inclusive deep 
inelastic scattering.

Effect of the spectator interaction can be explicitly calculated in the framework of the sum rule approach. 
Here the nucleon state is approximated by some current with appropriate quantum numbers, acting on the
vacuum. Then the problem is reduced to the calculation of the 4-point current correlator in vacuum.
It has poles at $p^2$ equal to the masses of physical states (nucleons) and discontinuity at the 
continuum. By means of dispersion relation it can be analitically continuated to space-like
region $p^2<0$, where it can be calculated as perturbative QCD series. Borel transformation by $p^2$
suppresses the contibution of the higher states and continuum and one obtains explicit expression 
for the structure functions in the intermediate $x$ region, where the sum rules are supposed to work.
Such a program was performed at the leading order in \cite{BI} for spin 1/2 nucleons.

The calculation of the $\alpha_s$ corrections to the nucleon structure functions will involve 3-loop graphs.
It is simpler, however, to begin from the pion case, since it has one loop less. The probabilistic interpetation, 
anomalous dimensions and the statement of \cite{BHMPS} do not rely on the hadron structure, 
so they are applicable to the pion as well. The subject of this paper is the calculation of 
the final state interaction in the pion. In the next section we briefly review the derivation of the 
pion structure functions at the leading order, performed in \cite{IO}, and then present original results.
Technical details are dropped to Appendix.

\section{Leading order pion structure functions}

One considers the 4-point current correlator:
\be
\Pi_{\mu\nu\alpha\beta}(p, q) \,=\,- \int dx \, dy \, dz \,e^{i(px + qy - pz)}
\, \langle	0 |  T j_{5\alpha}^\dagger (x) \, j_\mu^{em} (y) \, j_\nu^{em} (0) \, j_{5\beta}(z) | 0 \rangle
\label{4pcc}
\ee
where $j_{5\beta}={\bar u}\gamma_\beta \gamma_5 d$ is the pion axial current, 
$j_\mu^{em}=\sum_q e_q {\bar q}\gamma_\mu q$ electromagnetic one.
If momentum $p$ is close to the pion mass shell $p^2\approx m_\pi^2$, one may insert the
intermediate pion state, obtaining:
\be
\Pi_{\mu\nu\alpha\beta}(p, q) \,=\,{f_\pi^2 \, p_\alpha  p_\beta \over  (p^2-m_\pi^2)^2} \,
  \int dy \, e^{iqy} \, \langle \pi (p) | T j_\mu^{em}(y) \, j_\nu^{em}(0) | \pi(p)\rangle
\label{4pcc2}
\ee
where $f_\pi$ is the pion decay constant $\langle 0 | j_{5\beta}(0) |  \pi ( p ) \rangle = i f_\pi p_\beta $.
Authors of \cite{IO} took the incoming and outgoing pions with different momenta $p_1^2\ne p_2^2$
(but $(p_1q)=(p_2q)$). However, it is not necessary for our purposes: at the leading order results are equal. 
Moreover, the 2-loop calculation, performed in the next section, seems untractable in this case.
Correlator (\ref{4pcc2}) has discontinuity by the variable $s=(p+q)^2$:
\be
{\rm Disc}_s \Pi_{\mu\nu\alpha\beta} \,\equiv\, 
\Pi_{\mu\nu\alpha\beta}(s+i0) \,-\,\Pi_{\mu\nu\alpha\beta}(s-i0)\,=\,
8\pi {f_\pi^2 \, p_\alpha  p_\beta \over  (p^2-m_\pi^2)^2}\,W_{\mu\nu}
\ee
The tensor $W_{\mu\nu}$ consists of the pion structure functions.

In the quark basis the correlator (\ref{4pcc}) is given by the box diagram at the leading order:
\be
\begin{picture}(150,65)
\Photon(20,55)(50,50)25
\Photon(120,55)(90,50)25
\Line(50,10)(90,10)
\Line(50,50)(90,50)
\Line(50,12)(50,50)
\Line(90,12)(90,50)
\Line(20,5)(50,10)
\Line(20,7)(50,12)
\Line(120,5)(90,10)
\Line(120,7)(90,12)
\Text(11,10)[t]{$\pi$}
\Text(11,58)[t]{$\gamma^*$}
\Text(71,59)[t]{$u$}
\Text(44,33)[t]{$u$}
\Text(96,33)[t]{$u$}
\Text(71,8)[t]{$d$}
\end{picture}
\label{box_diag}
\ee
and the diagram with $u,d$ quarks exchanged. It is easy to show, that all other diagrams
(unlike the photon-photon scattering) are not essential in the Bjorken limit. 

Calculation of the diagram (\ref{box_diag}) gives the following imaginary part:
\be
{\rm Disc}_s \Pi_{\mu\nu\alpha\beta}(p,q)\,=\,-\,{3e_u^2\over \pi}\, {x(1-x)\over p^2}
 \,  p_\alpha p_\beta \left( \,-\,e_{\mu\nu}\, +\, {4x^2\over Q^2} \, \hat{P}_\mu \hat{P}_\nu \, \right)
\,+\, h.t.
\label{pi_lo}
\ee
$$
{\rm where} \qquad  e_{\mu\nu}\,=\,g_{\mu\nu}\,-\,{q_\mu q_\nu\over q^2}  \; , \qquad 
 \hat{P}_\mu\,=\,p_\mu-q_\mu {(pq)\over q^2} \, .
$$
Letters "$h.t.$" stand for the higher twist contributions, i.e. $\sim \ln{(-p^2)}$, 1, $p^2$ etc, 
small with respect to the leading term $1/p^2$ in the Bjorken limit $p^2\to 0$.

In the QCD sum rule approach one applies Borel operator to the variable $-p^2$
in order to suppress the contribution of the higher states and continuum. Then the following sum rule
can be obtained:
\bea 
 \hat{\cal B}_{M^2}{\rm Disc}_s \Pi_{\mu\nu\alpha\beta} 
 & = & {8\pi f_\pi^2\over M^4}\, e^{-m_\pi^2/M^2} \, p_\alpha p_\beta W_{\mu\nu} \nonumber \\
 &= & {3e_u^2\over \pi M^2}\, x(1-x)\,
 \,  p_\alpha p_\beta \left( \,-\,e_{\mu\nu}\, +\, {4x^2\over Q^2} \, \hat{P}_\mu \hat{P}_\nu \, \right)
\,+\, h.t.
\label{pi_sr}
\eea
The Borel mass should be taken \cite{IO}
\be
M^2 \,=\, 8\pi f_\pi^2
\ee
Then one extracts the $u$-quark distribution in the pion:
\be
f_u(x) \, =\, 6x(1-x)
\label{uq_distr}
\ee
In fact, the sum rules are not valid for $x$ close to 0 and 1, so exact value of $M^2$ as well as the
normalization integral 
\be
\int_0^1 dx \, f_u(x) \,=\, 1
\label{fx_norm}
\ee
are modified by various effects.  We will not pursue this point further, see \cite{IO} for details.

The sum rule (\ref{pi_sr}) demonstrates the following properties of the parton model:
\begin{itemize}
\item
Callan-Gross relation $F_2=2xF_1$.
\item
Bjorken scaling $F(x,Q^2)=F(x)$.
\item
Probabilistic interpretation of the function (\ref{uq_distr}) is supported by the normalization 
integral (\ref{fx_norm}).
\end{itemize}
The Bjorken scaling does not survive the $\alpha_s$-corrections. In particular, the leading order
evolution equation gives the $\ln{Q^2}$ term:
\be
\Delta f _{evol}\,= \,2\, {\alpha_s\over \pi} \ln{Q^2} \left[\, 1-x\,+\,4 \,(1-x)\, x\, \ln{(1-x)}\,-\,2\, (1-2 x)\, x\, \ln{x} \, \right]
\label{deltaf_evol}
\ee
Vacuum condensates also have their contributions, see \cite{IO}. In this paper we will pay attention
to perturbative calculation of the interaction with spectator.


\section{Gluon exchange with spectator}

In this section we will calculate the diagram 
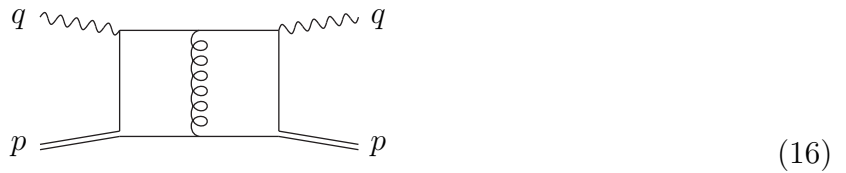
\be
\begin{picture}(150,65)
\Photon(10,55)(40,50)25
\Photon(130,55)(100,50)25
\Line(40,10)(100,10)
\Line(40,50)(100,50)
\Line(40,12)(40,50)
\Line(100,12)(100,50)
\Gluon(70,10)(70,50)36
\Line(10,5)(40,10)
\Line(10,7)(40,12)
\Line(130,5)(100,10)
\Line(130,7)(100,12)
\Text(2,10)[t]{$p$}
\Text(2,58)[t]{$q$}
\Text(138,10)[t]{$p$}
\Text(138,58)[t]{$q$}
\end{picture}
\label{2box_diag}
\ee
describing the interaction between the struck quark and spectator.  To avoid extra discontinuities,
the pion  momentum is taken to be space-like $p^2=-P^2<0$. 

We will consider its contribution to the tensor structure $p_\mu p_\nu p_\alpha p_\beta$, i.e. to the function
$F_2$. It it separated out by multiplying the diagram (\ref{2box_diag}) on the projector:
$$
{q^4\over d(d-2)\Delta^4} \left[ \, 3\Delta^2 \, e_{(\mu\nu} e_{\alpha\beta)}\, 
 +\,6(d+1)q^2 \Delta\, e_{(\mu\nu} \hat{P}_\alpha \hat{P}_{\beta)}
 \, + \, (d+1)(d+3) q^4\, \hat{P}_\mu \hat{P}_\nu \hat{P}_\alpha \hat{P}_\beta \, \right] 
$$
where $\Delta = (pq)^2-p^2q^2$. One can put the space-time dimention $d=4$ from the beginning
since, as explained below, only finite integrals will be important. Then the calculation is reduced to 
evaluation of some scalar integrals. 

Such 2-loop scalar integrals were considered in the calculation of the $\alpha_s^2$ anomalous dimensions 
of the structure functions \cite{2loopEV}. There it was sufficient to put $p^2=0$ from the very beginning,
since all IR poles cancel in the operator renormalization. 
In the sum rule approach  one may not accept such simplification, since the leading order integral (\ref{pi_lo})
already behaves as $1/p^2$. In dimensional regularization all the divergences  
have the pole $1/(d-4)$, regardless of their structure (UV or IR).  
For instance, it seems impossible to distinguish the terms 
$1/p^2$ and $\ln(-p^2)$ in such way. 
The first terms are interesting for us, while the later ones represent a higher twist correction.
Consequently, we have to keep $p^2$ finite.

The calculation is simpler in the Feynman gauge, since there are only 2 graphs of the leading twist:
\be
\begin{picture}(330,70)
\Line(30,15)(110,15)
\DashLine(30,55)(110,55)4
\Line(30,15)(30,55)
\Line(110,15)(110,55)
\Line(70,15)(70,55)
\Line(10,7)(30,15)
\Line(10,62)(30,55)
\Line(130,7)(110,15)
\Line(130,62)(110,55)
\Text(2,10)[t]{$p$}
\Text(2,63)[t]{$q$}
\Text(23,40)[t]{$k$}
\Text(117,40)[t]{$l$}
\Text(49,12)[t]{$k-p$}
\Text(91,12)[t]{$l-p$}
\Text(49,69)[t]{$k+q$}
\Text(91,69)[t]{$l+q$}
\Text(84,40)[t]{$k-l$}
\Line(210,15)(270,15)
\DashLine(210,55)(270,55)4
\Line(210,15)(210,55)
\Line(270,15)(270,55)
\Line(245,15)(210,55)
\Line(190,7)(210,15)
\Line(190,62)(210,55)
\Line(290,7)(270,15)
\Line(290,62)(270,55)
\end{picture}
\label{lt_diags}
\ee
Indeed, consider the left graph of eq (\ref{lt_diags}). Let us choose a frame $p=(0,P,0_\bot)$ and
take a part of the phase space with the integration momenta $k,l\sim p$. The volume 
of this region is $\int d^4k\, d^4 l \, \sim p^8$. Each propagator, depicted by the solid line on the graph,
brings the factor $1/p^2$ here, while the dashed line brings $1/q^2$. 
Consequently the left graph has the order $p^8/(p^2)^5(q^2)^2=1/(p^2q^2)$.
The second graph  of the leading twist, the right one in eq (\ref{lt_diags}), is obtained 
by removing single dash line. (If both dash lines are removed, the integral becomes real, since no intermediate
state can be inserted.) But if any solid line is removed, the twist  increases and such diagram can be ignored.

However, the most complicated 2-box graph of (\ref{lt_diags}) turns out to be real:
\be
{\rm Disc}_s  \int_{k,l} {1 \over k^2\, l^2 \, (k-l)^2\, (k-p)^2 \, (l-p)^2\,(k+q)^2\, (l+q)^2}\,=\,0
\label{2box0}
\ee
We do not know a simple way to prove this result, so we put (rather cumbersome) details of the calculation
in Appendix. In the evaluation of (\ref{2box_diag}) one encounters 2 integrals of the 
box-triangle leading twist topology:
\bea
J_1 & = & {\rm Disc}_s \int_{k,l} {1 \over k^2\, l^2 \, (k-l)^2\, (k-p)^2 \, (l-p)^2\, (l+q)^2} \label{int_j1} \\
J_2 & = & {\rm Disc}_s \int_{k,l} {2(kq)+q^2 \over k^2\, l^2 \, (k-l)^2\, (k-p)^2 \, (l-p)^2\, (l+q)^2} \label{int_j2}
\eea
Exact results for these integrals (arbitrary $p^2$, $q^2$, $s$) are available in Appendix (\ref{tot_int2}),
 (\ref{tot_int22}).

The contribution of the diagram (\ref{2box_diag}) to the structure $p_\mu p_\nu p_\alpha p_\beta$
of the correlator (\ref{4pcc}) discontinuity is:
\be
 2^{11}\pi e_u^2\alpha_s  \,ix (1-x) \, \left[ \, -\, (1-2x)\, J_1\,+ \,x\, Q^{-2}\,J_2\,  \right]  +\,h.t.
\ee
Substituting it into the sum rule (\ref{pi_sr}), we obtain the following correction to the
quark distribution function (\ref{uq_distr}):
\bea
\Delta f(x) &= & {8\alpha_s\over \pi} x(1-x) \Biggl\{ \,3(1-2x)\left[\,{\rm Li}_3\!\left(1-{1\over x}\right) - \zeta_3 \right]
   - \,(1-2x)\,\ln\!{\left({1\over x}-1\right)} \nonumber \\
 & & 
\times \left[ \, {\rm Li}_2\!\left(1-{1\over x}\right) -{\pi^2\over 6}\right]
 - \,2\,{\rm Li}_2\!\left(1-{1\over x}\right) + \ln{x}\, \ln\!{\left({1\over x}-1\right)} - {\pi^2\over 6}\, \Biggr\} 
\label{deltaf}
\eea
Notice, that 
\be
\int_0^1 \Delta f(x) \, dx \,=\, 0 
\ee
So this correction does not change the structure function normalization (\ref{fx_norm}).

\section{Conclusion}

We have calculated a leading-twist $\alpha_s$ correction to the pion structure functions (\ref{deltaf})
due to the final state gluon exchange between the struck quark and spectator in the Feynman gauge.
We certainly do not claim, that it is not cancelled by the initial state interaction  or
does not vanish in some specific gauge. Although the arguments, based on the operator expansion
of the 2-point current product are of no doubt, they are less obvious being applied to the 
4-point current correlator, considered in the sum rule approach. The spectator interaction, which occurs immediately
before or after the virtual photon absorption, may give rise to a leading twist effect beyond the series
(\ref{tw_ser}).

The initial state diagram (\ref{qcdint_diag}b) involves 
more scalar graphs, which require additional cumbersome calculations to perform.
If it does not cancel the final state interaction in the sum, this will support the 
statement of \cite{BHMPS}, and probabilistic interpretation of the structure functions 
should be reconsidered. But even if it does, the calculation of the diargams 
(\ref{qcdint_diag}a) with only the struck quark involved would also give a new 
information: its infinite part must reproduce the known $\ln{Q^2}$ contribution 
(\ref{deltaf_evol}), determined by the evolution, while the finite part would be a new correction. 
These diagrams altogether will provide us with refined pion
structure functions to be fitted with experimental data.

\section*{Acknowledgment}

Author thanks B.L. Ioffe and A.G. Oganesian for discussions. 
This work was partially supported by CRDF Cooperative Grant Prorgam RUP2-2621-MO-04
 and  RFBR grant 03-02-16209.


\section*{Appendix}

\setcounter{equation}{0}
\def\theequation{A\arabic{equation}}

Below we describe the calculation of the two-loop integrals (\ref{2box0}) and (\ref{int_j1}), (\ref{int_j2}).

{\bf Notations and conventions.} Vecotors $p$, $q$ are taken to be space-like, while $p+q$ is
the time-like one:
\be
p^2\,=\,-P^2\,<\,0  \; , \qquad q^2\,=\,-Q^2\,<\,0  \; , \qquad s\,=\,(p+q)^2\, >\, 0  \; .
\ee
It is convenient to perform the calculation in the center of mass frame $p+q=(\sqrt{s},0)$. One
may choose
\be
p\,=\,(\sqrt{\omega^2-P^2},0,0,\omega)  \; , \qquad 
q\,=\,(\sqrt{\omega^2-Q^2},0,0,-\omega)
\label{pqframe}
\ee
where 
\be
\omega=\sqrt{\nu^2-P^2 Q^2\over s}  \; , \qquad \nu=(pq) \, .
\ee
In the calculation of the 2-particle cut we use the 2-particle phase volume:
\bea
d\Gamma_2 & \equiv & \int {d^3k\over (2\pi)^3 2k^0}\,{d^3k_2\over (2\pi)^3 2k_2^0}\,
(2\pi)^4 \delta^4 (p+q-k-k_2)  \nonumber \\
 &=& {1\over 16\pi \, \omega \sqrt{s}} \, \int_{T_-}^{T_+}\! dT
\label{dgam2}
\eea
where $k^2=k_2^2=0$ and
\be
T\,=\,-(p-k)^2  \; , \qquad T_\pm \,=\,\nu \pm \omega \sqrt{s}
\ee
In (\ref{dgam2}) we integrated out azimuthal angle $\phi$ in $xy$ plane, since 
in the frame (\ref{pqframe}) the amplitudes do not depend of it.

The 3-particle phase volume is:
\bea
d\Gamma_3 & = & \int {d^3 k_1\over (2\pi)^3\, 2k_1^0} \,
 {d^3 k_2\over\, (2\pi)^3 2k_2^0}\,{d^3 k_3\over (2\pi)^3\, 2k_3^0} \,(2\pi)^4\, \delta^4(p+q-k_1-k_2-k_3) 
 \nonumber \\
 & =  & {1\over (2\pi)^5} \int {d^3 k_1\over 2 k_1^0}\, {d^3 k_2\over 2 k_2^0}  \,
\delta[(p+q-k_1-k_2)^2-\lambda^2]\, \theta(p^0+q^0-k_1^0-k_2^0)
\label{dgam3}
\eea
where $k_1^2=k_2^2=0$, but $k_3^2=\lambda^2$, the mass $\lambda$ is necessary to regularize the IR
divergencies. In spherical coordinates
\bea
d\Gamma_3& = & {1\over (2\pi)^5} \cdot {1\over 8} \int_0^{\sqrt{s}}\! dk_1 \int_0^{\sqrt{s}-k_1}\! dk_2
\int_0^\pi\! d\theta_1 \int_0^\pi\! d\theta_2 \int_0^{2\pi}\! d\phi_1 \int_0^{2\pi}\! d\phi_2 \nonumber \\
 & & \times \,
\delta\left[\, {\cos\alpha - \cos\theta_1 \cos\theta_2  \over \sin\theta_1\sin\theta_2 }
\, -\, \cos(\phi_1-\phi_2)\, \right]
\label{dgam3_2}
\eea
where we denoted
$$
\cos\alpha\,=\,{s-2\sqrt{s}(k_1+k_2) +2k_1k_2-\lambda^2 \over 2 k_1 k_2}  \, .
$$
For a given $\theta_1$ the condition $\cos^2(\phi_1-\phi_2)<1$ sets the following restriction on 
the integration range:
$$
|\cos{\alpha}|<1  \; , \qquad \cos{(\theta_1+\alpha)} < \cos\theta_2 < \cos{(\theta_1-\alpha)}
$$
It is convenient to introduce new variable $\xi$ instead of $\theta_2$ and $t_{1,2}$ instead 
of $k_{1,2}$:
\be
\cos\theta_2\,=\,\cos\theta_1 \cos\alpha + \sin\theta_1 \sin\alpha\cos\xi  \; , \qquad 
k_i\,=\,{\sqrt{s}\over 2}(1-t_i)
\ee
The $\delta$-function is excluded by integrating out $\phi_2$. Since the amplitudes are independent of the 
angle $\phi_1$, we integrate it also. Then the 3-particle  phase  
volume (\ref{dgam3_2}) takes the most convenient form:
\be
d\Gamma_3 \,=\,  {1\over (2\pi)^4} \cdot {s\over 16} \int_{\lambda^2/s}^1 \! dt_1 
 \int_{\lambda^2/(st_1)}^{1-t_1+\lambda^2/s} \! dt_2 
 \int_0^\pi \sin \theta_1 \, d\theta_1 \int_0^\pi d\xi 
\ee

{\bf 2-box diagram} consists of two cuts
\be
\begin{picture}(140,65)
\Line(30,15)(110,15)
\Line(30,55)(110,55)
\Line(30,15)(30,55)
\Line(110,15)(110,55)
\Line(70,15)(70,55)
\Line(10,7)(30,15)
\Line(10,62)(30,55)
\Line(130,7)(110,15)
\Line(130,62)(110,55)
\DashLine(55,7)(85,62)4
\DashLine(95,7)(95,62)4
\Text(50,3)[t]{$Cut_3$}
\Text(95,3)[t]{$Cut_2$}
\Text(62,45)[t]{$\lambda^2$}
\Text(2,10)[t]{$p$}
\Text(2,64)[t]{$q$}
\end{picture}
\label{2box_d}
\ee
and symmetric ones. Since all internal lines are massless, both cuts are IR divergent. A common way
to regularize it is the introduction of small  "photon" mass $\lambda^2$. To evaluate the 2-particle
cut, one needs to know the 1-loop box integral:
\be
I_{\Box}\,=\,\int_l{1\over l^2 \,(l-p)^2\, (l+q)^2\, [(l-p+k)^2-\lambda^2]}
\ee
The Feynman representation of the propagator product yelds the following 3-fold integral:
$$
I_{\Box} \, = \,  {i \over 16\pi^2} \int_0^1 dx \int_0^x dy  \int_0^y dz\, \times \hspace{87mm}
$$ 
\be
\label{box_int1} 
\times \left[ Q^2(x-y)(1-x) + Tz(1-x) + P^2(y-z)(1-x) -s(x-y)(y-z) +\lambda^2 z \right]^{-2} 
\ee
This integral is real for $s<0$, since the denominator is never zero in this case. 
So we calculate it for negative $s$ and then analytically continuate to positive region.
The result is:
\be
I_{\Box}=-{i\over 16\pi^2 sT} \left[\, {\rm Li}_2\left( 1-{T\over P^2} \right) +
\, {\rm Li}_2\left( 1-{T\over Q^2} \right)  +\, {1\over 2}\, \ln^2\!{ -sT^2 \over \lambda^2 P^2 Q^2}
 \,+\,{\pi^2\over 3}\, \right]
\label{box_int2}
\ee
Then one calculates the contribution of both 2-particle cuts:
\bea 
2 \, Cut_2 & = & \int   d\Gamma_2  {i \over (p-k)^2} \left\{ \left. I_{\Box}\right|_{s+i0} +
 \left. I_{\Box}\right|_{s-i0} \right\}  \nonumber \\
 &= & -\,{1\over 128 \pi^3 sP^2Q^2} \Biggl\{ \ln^2\!{\lambda^2\over s} - 4\ln{\lambda^2\over s}
 +8-{\pi^2\over 3} +3\ln^2\!{T_-\over PQ} - \ln^2\!{P\over Q}   \nonumber \\
 & & 
+ {4\nu\over \omega\sqrt{s}}\ln{T_-\over PQ} \left(2-\ln{\lambda^2\over s} \right)  
- {P^2-Q^2\over \omega\sqrt{s}}\ln{P\over Q} \ln{T_-\over PQ} \nonumber \\
 & & 
+ {\sqrt{s}\over \omega}\left[ {\rm Li}_2 \left( 1- {P^2\over T_-}\right) + {\rm Li}_2 \left( 1- {Q^2\over T_-}\right)
 + {1\over 2}\ln^2\!{T_-\over PQ}+ {1\over 2}\ln^2\!{P\over Q} \right]
 \Biggr\}
\label{2box_cut2}
\eea
The 3-particle cut is given by the integral:
\bea
Cut_3 & = & \int{d\Gamma_3\over (p+q-k_1)^2(p+q-k_2)^2(p-k_1)^2(q-k_2)^2}  \\
 &= & \int{d\Gamma_3 \,(s-2\sqrt{s} k_1)^{-1}(s-2\sqrt{s}k_2)^{-1}\over 
 [-P^2-2k_1(\sqrt{\omega^2-P^2}-\omega \cos\theta_1)]
[-Q^2-2k_2(\sqrt{\omega^2-Q^2}+\omega \cos\theta_2)]}\nonumber
\eea
Integration over the angles yields:
\bea
Cut_3 & = & {1\over 64\pi^3 s} \cdot {1\over 4\omega\sqrt{s}} \int_{\lambda^2/s}^1 {dt_1\over t_1}
 \int_{\lambda^2/(st_1)}^{1-t_1+\lambda^2/s} {dt_2\over t_2}  \label{3cut_intk} \\
 & &\times {1\over st_1t_2+P^2t_1+Q^2t_2}\ln{(1-t_1-t_2)P^2Q^2+T_+(st_1t_2+P^2t_1+Q^2t_2)\over 
(1-t_1-t_2)P^2Q^2+T_-(st_1t_2+P^2t_1+Q^2t_2)} \nonumber
\eea
This integral is IR divergent at $t_{1,2}\to 0$ (or $k_{1,2}\to {\sqrt{s}\over 2}$). The expresion
in the second line of  (\ref{3cut_intk}) is regular there, so we put $\lambda^2=0$ everywhere except
the integration limits; account of the terms neglected would lead to power corrections in $\lambda$,
but not logarithmic ones.
Evaluation of (\ref{3cut_intk}) yields exactly $-1/2$ of equation (\ref{2box_cut2}), so
$$
2\, Cut_2 \, + \,2 \, Cut_3\,=\,0
$$
which proves the equation (\ref{2box0}).

{\bf Box-triangle diagram} 
\be
\begin{picture}(120,65)
\Line(30,15)(90,15)
\Line(30,55)(90,55)
\Line(30,15)(30,55)
\Line(90,15)(90,55)
\Line(65,15)(30,55)
\Line(10,7)(30,15)
\Line(10,62)(30,55)
\Line(110,7)(90,15)
\Line(110,62)(90,55)
\DashLine(45,7)(67,62)4
\DashLine(78,7)(78,62)4
\Text(40,3)[t]{$Cut_3$}
\Text(80,3)[t]{$Cut_2$}
\Text(43,34)[t]{$\lambda^2$}
\Text(2,10)[t]{$p$}
\Text(2,64)[t]{$q$}
\end{picture}
\label{bt_diag}
\ee
is calculated in similar way. The triangle 1-loop integral is:
\bea
I_\Delta & = & \int_l {1\over l^2\, (l+p)^2 \, [ (l+k)^2-\lambda^2 ] } \nonumber \\
 &= &  -{i\over 16\pi^2 (T-P^2)}\left[ \,{\rm Li}_2\left( 1- {P^2\over T} \right) 
  + \, \ln{T\over \lambda^2}\ln{T\over P^2} \,  \right]
\label{tr_loop}
\eea
The 2-particle cut is equal to:
\bea
Cut_2 & = & \int d\Gamma_2 {i \over  (k-p)^2} \,i^3\, I_\Delta  \nonumber \\
 &= & 
{i\over 256\pi^3 \omega\sqrt{s}P^2}\Biggl[ \,2\,{\rm Li}_3(1-y)\, +\, 3\,{\rm Li}_3(1-1/y)\,-\,2\,\ln{y}\,{\rm Li}_2(1-y)
  \nonumber \\
 & &  \left. \left. \,-\, {2\over 3}\,\ln^3\!y \,
 + \ln{P^2\over \lambda^2} \, {\rm Li}_2(1-1/y) \right]\right|_{y_-}^{y_+}  \; , \qquad y={T\over P^2}
\label{cut2_tr_b}
\eea
The 3-particle cut is given by the same integral as (\ref{3cut_intk}), but without $t_1$ in denominator:
\bea
Cut_3 & = & \int {i^3\,  d\Gamma_3\over (p+q-k_2)^2 \,  (p-k_1)^2 \, (q-k_2)^2} \nonumber \\
 & = &  {i \over 256\pi^3 \omega\sqrt{s} P^2} \, \Biggl\{ \,
\left[\ln{y_-\over y_+} - \ln{s\over \lambda^2} \right] {\rm Li}_2(1-1/y_+)\nonumber \\
 & & 
 - \,2\,{\rm Li}_3( 1-y_+)\,-{1\over 3}\ln^3y_+ \,-\,(y_-\leftrightarrow y_+) \Biggr\}  
\label{cut3_tr_b}
\eea
In the calculation we used the following properties of the function ${\rm Li}_3$:
$$
{\rm Li}_3\!\left(1 -z\right) +
{\rm Li}_3\!\left( 1-1/z \right) + {\rm Li}_3(1/z) =
\zeta_3-{\pi^2\over 6}\ln{z}+{1\over 3}\ln^3\!{z}-{1\over 2}\ln\!{(z-1)}\, \ln^2\!{z}
$$
$$
{\rm Li}_3(-z)\,-\,{\rm Li}_3(-1/z)\,=\,-\,{\pi^2\over 6}\ln{z}\,-\,{1\over 6}\ln^3\!{z}
$$
Summing up (\ref{cut2_tr_b}) and (\ref{cut3_tr_b}), one obtains the integral (\ref{int_j1}):
\bea
J_1 & = & -\,{i\over 256\pi^3 \omega\sqrt{s} P^2}\left[ \,3\,{\rm Li}_3\!\left( 1-{P^2\over T_+} \right)
 - \,3\,{\rm Li}_3\!\left( 1-{P^2\over T_-} \right) \right. \nonumber \\
 & & \hspace{15mm} \left. +\, \ln{Q^2\over s}\, {\rm Li}_2\!\left( 1-{P^2\over T_+} \right)
 -\, \ln{Q^2\over s}\, {\rm Li}_2\!\left( 1-{P^2\over T_-} \right) \right]
\label{tot_int2}
\eea
The intergal (\ref{int_j2}) is calculated in similar way:
\bea
J_2 &  = & -\,{i\over 256\pi^3\omega \sqrt{s} P^2}
\left[ \,(T_- - 2T_+) \, {\rm Li}_2\!\left( 1-{P^2\over T_-} \right) 
\, -\,(T_+ - 2T_-) \, {\rm Li}_2\!\left( 1-{P^2\over T_+} \right) \right.  \nonumber \\
 &  & 
\hspace{30mm} 
 \left.  -\, T_+ \ln{T_-\over P^2} \ln{Q^2\over s}\, +\, T_- \ln{T_+\over P^2} \ln{Q^2\over s} \, \right]
\label{tot_int22}
\eea
In the Bjorken limit $P^2\to 0$ and
\be
\omega\sqrt{s}\approx \nu ={Q^2\over 2x}  \; , \qquad 
 s\approx Q^2 {1-x\over x}  \; , \qquad 
 T_+\approx {Q^2\over x}  \; , \qquad T_-\approx xP^2 \, . 
\ee
Notice cancellation of the terms $\sim \ln^n \! P^2/P^2$ in (\ref{tot_int2}), (\ref{tot_int22});
 they would violate Bjorken scaling.


\begin{thebibliography}{99}
\bibitem{J}
R.L. Jaffe, hep-ph/9602236
\bibitem{2loopEV}
A. Devoto, D.W, Duke, J.D. Kimel, G.A. Sowel, Phys. Rev. D30 (1984) 541.
D.I. Kazakov, A.V. Kotikov, Nucl. Phys. B307 (1988) 721; e. idid. B345 (1990) 299.
W.L. van Neerven, E.B. Zijlstra, Phys. Lett. B272 (1991) 127; B273 (1991) 476; B297 (1992) 377.
S. Moch, J.A.M. Vermaseren, Nucl. Phys. B573 (2000) 853.
\bibitem{3loopEV}
S. Moch, J.A.M. Vermaseren, A. Vogt, Nucl. Phys. B688 (2004) 101; B691 (2004) 129; hep-ph/0504242
\bibitem{S}
A.V. Smilga, Phys. Lett. B 86 (1979) 75
\bibitem{JMT}
R.L. Jaffe, Nucl. Phys. B229 (1983) 205.
P.J. Mulders, R.D. Tangerman, Nucl. Phys. B461 (1996) 197; e.ibid. B484 (1997) 538
\bibitem{BHMPS}
S. J. Brodsky, P. Hoyer, N. Marchal, S. Peigne, F. Sanino, Phys. Rev. D65 (2002) 114025
\bibitem{JY}
X. Ji, F. Yuan, Phys. Lett. B543 (2002) 66.
A.V. Balitsky, X. Ji, F. Yuan, Nucl. Phys. B656 (2003) 165
\bibitem{Betal}
S. J. Brodsky, D.S. Hwang, I. Schmidt, Phys. Lett. B530 (2002) 99; Nucl. Phys. B 642 (2002) 344.
D. Boer, S.J. Brodsky, D.S. Hwang, Phys. Rev. D67 (2003) 0504003
\bibitem{BI}
V.M. Belyaev, B.L. Ioffe, Nucl. Phys. B310 (1988) 548
\bibitem{IO}
B.L. Ioffe, A.G. Oganesian, Eur. Phys. J. C 13, 485 (2000)
\end{thebibliography}
\end{document}